\documentclass{article}

\usepackage[final]{neurips_2019}

\usepackage[utf8]{inputenc}
\usepackage[T1]{fontenc}
\usepackage{url}
\usepackage{booktabs}
\usepackage{amsfonts}
\usepackage{nicefrac}
\usepackage{microtype}
\usepackage{graphicx}
\usepackage{xcolor}
\usepackage{lipsum}
\usepackage{amsmath,amssymb}
\usepackage{hyperref}

\title{
  Unraveling the Dynamics of SPY Trading Volumes:\\
  A Comprehensive Analysis of Daily and Intraday Liquidity Trends \\
  \vspace{1em}
}

\author{
  Ananya Krishnan \\
  Department of Statistics \\
  Stanford University \\
  \texttt{ananyakr@stanford.edu} \\
   \And
   Martin Pollack \\
   Department of Statistics\\
   Stanford University \\
   \texttt{pollackm@stanford.edu}
   \And
      Alma Cooper \\
   Department of Statistics\\
   Stanford University \\
   \texttt{alma1@stanford.edu} \\
}

\begin{document}

\maketitle
\thispagestyle{empty}

\begin{abstract}
  In this project, we investigate the accuracy of forecasting intraday and daily trading volume of the exchange-traded fund SPY. The ability to forecast volume over varying time intervals with high accuracy is a critical element to many trading strategies. After performing exploratory data analysis on intraday and daily SPY data we identify three methods for our analysis: ARIMA and ARIMAX models, with or without seasonality, as well as a Frequency Domain Process Representation. To evaluate predictive power of our models, we use mean squared error, mean absolute percentage error, and volume weighted average price (VWAP) tracking error. All models for both intraday and daily data output strong VWAP predictions in comparison to the VWAP estimates produced by naive baseline methodologies. In both cases volume is most accurately forecasted using ARIMA models with exogenous variables in the form of technical indicators, with intraday incorporating a seasonal component and daily not.
  
\end{abstract}



\section{Introduction}

This study examines intraday and daily traded volume of SPY, an exchange-traded fund (ETF) tracking the S\&P 500 index and managed by State Street. As the most traded ETF globally, SPY is widely held by major asset managers both in the United States and worldwide. While each institution has different trading strategies, almost all strategies require a decent understanding of the evolution of trading volume over time. Buy-side institutions must carefully plan and time their trades to avoid significantly impacting the market, revealing their identities, and taking on excessive transaction costs. Sell-side institutions require knowledge of trading volume to make markets efficiently and numerical forecasts to implement any strategies related to volume, a common set being those that track some form of volume weighted average price (VWAP). This study aims to analyze SPY's liquidity picture over time and understand whether trading volumes can be accurately forecasted at the hourly and daily levels.

It is common knowledge within trading communities that, within a given day, most assets tend to see a spike in liquidity near market close, a smaller spike in the hours after market open, and a general lull midday. While there are some concrete reasons for this (e.g. trades that track indices will likely trade near market close, investors react to overnight news in the morning, and traders are more likely to be off-desk for lunch during the early afternoon hours), a large portion is due to happenstance. The game-theoretic aspect of trading is wanting to trade when everyone else is trading, and through time market participants have converged to trade more at these times of the day. However, these statements are all qualitative observations and are not directly useful for institutions looking to plan out their trading strategies. As such, our first goal will be to quantify and forecast the intraday liquidity picture.

While less studied than intraday volume, volume in the markets across consecutive days is anecdotally known to be highly correlated. To provide some intuition behind this, under the simplifying assumption that there is a significant and positive correlation between price volatility and trading volume, consider a day where everyone is dumping shares (e.g. after an earnings release where the company failed to beat estimates). At a first-order level, market participants may see the prices crashing and shift to a more conservative loss-minimization strategy, leading them to sell more than they otherwise would have. Additionally, rapid price drops may trigger stop-losses or limit trades, leading to even larger amounts of market volume being pushed through exchanges. On days following large drops, as more information continues to disseminate, there are typically either rebounds in price movement or additional drops, both of these lesser in magnitude than the day of the news release, but with above-average volatility and thus, above-average volumes. As more days pass from the initial drop, the news becomes more fully priced into the markets, leading price volatility and volume to return to the typical range. Again, these are anecdotal observations and are not directly useful inputs into trading strategies, so our second goal will be to quantify the evolution of day-over-day liquidity.

In choosing our models, we focus almost exclusively on predictive power. This is because traders care mostly about having an accurate estimate of future volumes, and they are less concerned about how well a model fits the data or whether the data is actually generated by a certain model. Using cross validation with metrics Mean Square Error (MSE) and Mean Absolute Percentage Error (MAPE), we find that the best predictive model for intraday volume is SARIMAX. We find a similar result for daily volume, but without the presence of seasonality. We then further evaluate the performance of these best cross validation models using VWAP tracking error (see Equation \ref{vwap}).

\begin{equation}
     Err_{VWAP} = \frac{VWAP_{pred} - VWAP_{actual}}{VWAP_{actual}}
     \label{vwap}
\end{equation}

We find that all of our models lead to much superior VWAP predictions compared to naive strategies to estimate VWAP.

\section{Related Work}
Recent literature includes many examples of time series models being used to predict values related to stocks. However, a majority of these models focus on stock price or returns information. Stock volume is only investigated in the context of how it is related to price or returns, and this is done mostly in the setting of causal analysis. For example, Chiang et al. \cite{Chiang} focus on the relationship between trading volume and return volatility, using a simple stationary de-trended model to make predictions and conduct linear Granger causality tests. Similarly, Batrinca et al. \cite{Batrinca} employ a causal analysis to understand which factors affect trading volume in European markets, finding that historical stock price, price asymmetry, and price-volume relation asymmetry are the most important drivers of volume. To undergo this analysis, Batrinca et al. \cite{Batrinca} use both autoregressive and moving average models, running statistical tests on model coefficients to understand statistical significance of causal relationships. Then in So et al. \cite{So} a nonlinear threshold time-series model with generalized autoregressive conditional heteroscedastic disturbances and Markov chain Monte Carlo procedures are used to show how trading volume has a large effect on market returns. Lastly, Liu et al. \cite{Liu} use trading volume as a covariate along with stock price volatility to estimate future stock prices with a Long-Short Term Memory (LSTM) neural architecture. None of these models of increasing complexity make trading volume prediction a priority. Thus, as getting accurate trading volume predictions are vital for traders in practice, we focus on building the best predictive volume model rather than doing another causal analysis.

Given all the causal analysis papers focusing on the strong relationship between trading volume and stock price, it seems important to incorporate stock price data into our volume predictions. In the literature, the way stock prices are usually introduced as covariates to understand market behaviours is through technical indicators, values calculated from historical stock prices that help investors find patterns in the market and anticipate future movements. For example, Oriani et al. \cite{Oriani} employ technical indicators in predicting stock closing values. They consider twelve different technical indicators and show that to varying degrees they aid in the prediction of closing values using artificial neural network models. Batrinca et al. \cite{Batrinca} highlight that technical indicators are a significant driver of trading in volume in Europe. Our report builds on this previous work, showing that technical indicators can be used to accurately forecast trading volume.

\section{Data}
For intraday analysis, we pull hourly trading volume data from the Alpaca API. The data provided by the API is pulled directly from the Consolidated Tape Association (CTA) and Unlisted Trading Privileges stream (UTP), both of which are directly administered by the NYSE and NASDAQ exchanges respectively. Alpaca provides us with the last 8 years worth of hourly data, giving us over 33,000 observations. We begin by filtering our dataset to observations in 2024. The main reason for this is that intraday trading volume can fluctuate with the economic environment and as more instruments become available to gain similar exposure, so we want to choose a relatively short and recent period of time that provides us a good indication of the current market conditions. We then remove all weekends, trading holidays, pre-market and after-hours data and adjust our data for daylight savings, leaving us with around 1,500 observations.

To examine SPY's day-over-day liquidity picture, we use Yahoo Finance to collect daily trading volume, with the original data provider being Commodity Systems Inc. The data encapsulates information from the fund's launch on January 29, 1993 to May 25, 2024, leaving us with over 7,000 trading days worth of data. Again, we want to restrict our analysis to more recent data fearing that going too far into the past would yield data created under different economic conditions. In times of uncertainty, like during the COVID-19 pandemic, market volatility tends to spike significantly. This usually leads market participants to shift into more conservative strategies, impacting market dynamics in ways that are related to the specifics of the crisis. In periods like these, algorithmic trading usually becomes more manual in nature as traders re-evaluate the evolution of day-to-day market conditions, changing the nature and quantity of trades placed on any given day. Given trading volume forecasts are often inputs into trading algorithms, we choose not to model periods with exceptional market volatility. Thus, for daily volume data we restrict ourselves to a post-pandemic time frame, January 1, 2023 to May 25, 2024.

\section{Methods}
\subsection{ARIMA/SARIMA}
Our methods focus mostly on finding the best model to predict SPY volumes either at a level of granularity of intraday or daily.

Through our exploratory data analysis (EDA) which we explain in our Experiments section, we find strong seasonality components in the intraday and daily data, prompting us to start our analysis with a (Seasonal) ARIMA model ((S)ARIMA). By using this model, we assume that after applying differencing $d$ times and potentially seasonal differencing $D$ times to our volume time series we will be left with a multiplicative seasonal ARMA model, which in turn is a combination of ARMA and seasonal ARMA. Concretely, this means we are modeling our time series $Y_t$ as
$$
\Phi(B^s)\phi(B)\triangledown_s^D\triangledown^d Y_t=\delta+\Theta(B^s)\theta(B)Z_t,
$$
where $\Phi(B^s)$ is the seasonal AR operator, $\phi(B)$ the AR operator, $\Theta(B^s)$ the seasonal MA operator, $\theta(B)$ the MA operator, $\triangledown_s^D$ the seasonal differencing operator, $\triangledown^d$ the differencing operator, and $\delta$ a parameter related to drift. The seasonal operators are removed for ARIMA models. We also suppose that $Z_t$ is a white noise process. (S)ARIMA models of various orders were fit and compared using average MSE and MAPE metrics obtained from cross validation.

\subsection{SARIMAX}
However, in reality traders do not just use past volume data to predict future volumes. Additional covariates, or exogenous variables, are also commonly used to create robust estimates as seen in Batrinca et al. \cite{Batrinca} and Oriani et al. \cite{Oriani}. For example, stock prices can affect volume. If stock prices suddenly drop, stop-losses and limit orders automatically get executed and people may also be incentivized to manually trade into the volatility, boosting trading volume. As a result, we decide to consider a second model: Seasonal ARIMA with Exogenous Regressors (SARIMAX). The formula for this model is identical to SARIMA with the addition of the term $\sum_{i=1}^p\beta_i X_{i_t}$ where $p$ is the number of covariates added to the model and $\beta_i$ is the coefficient of $X_{i_t}$, or the value of the $i$th covariate at timestep $t$. This term allows the next volume value to not just depend on moving average, autoregressive, differencing, and seasonal terms, but rather also values of our covariates. See Equation \ref{eq:SARIMAXModel} in the Appendix for the full formulation.

This leads to the question: which covariates should we use? As mentioned above, in the literature technical indicators are used to give further context on stock predictions. We consider adding six different technical indicators to our SARIMAX model: Average Directional Index (ADI), Exponential Moving Average (EMA), Momentum (MOM), Rate of Change (ROC), Relative Strength Index (RSI), and Williams \%R (WPR). Descriptions of these technical indicators can be found in Appendix Table \ref{tab:TechnicalIndicators}. We then decide which covariates to include in our final SARIMAX model using a forward stepwise regression approach along with cross validation. We start with the null model and iteratively add covariates one at a time and select the model that maximizes average MSE and average MAPE values. More specifically, we first find the covariate that on its own led to best predictions using cross validation, where we iterate over our dataset to fit a model with certain data and predict the volume for the following trading day. Then we fix this best covariate and searched for the second covariate which led to the biggest improvement, continuing this process of adding covariates to our model until overfitting occurs.

\subsection{Frequency Domain Process Representation (FDPR)}
Lastly, we examine the data from a frequency domain approach. We can approximate any stationary time series as a sum of sines and cosines of certain frequencies. This is, of course, provided we sum enough of these sinusoidal functions together, something which is decided by the hyperparameter $m$. Thus, our approach was to find the optimal value of $m$ to create the frequency domain model with the best predictive power.

In order to accomplish this, we use a form of cross validation with the following steps. Fixing the value of $m$ ahead of time, for each timestep in our testing period we use the periodogram of the training period to calculate the $m$ most important Fourier frequencies. Then we use linear regression to estimate the amplitudes of the sines and cosines for these Fourier frequencies, which is reasonable given that the spectral representation of a process assumes the amplitudes have mean zero and constant variance. Note that not all assumptions of linear regression may have been met by our data; however, we only care about predictive power for this analysis. Thus, violating model assumptions is of less importance to us. With these Fourier frequencies and amplitudes we then are able to sum sines and cosines which yielded a function, a function with which we predicted the volume for the testing period. We then use average MSE and MAPE over all testing periods to choose the value of $m$ which yielded the best predictions.

The three models described here are used to get predictions for intraday and daily trading volumes. We compare their MSE, MAPE, and VWAP error values. We also qualitatively inspect the graphs of predicted values to make our final decisions regarding the best models. Note that all implementations were created and coded by the members of this team. This includes data prepossessing, exploratory data analysis, and cross validation. Only R functions for fitting ARIMA models like \texttt{auto.arima()} and for linear regression models like \texttt{lm()} were leveraged.

\section{Experiments}


\subsection{Intraday}

We begin by discussing our approach to examining intraday volume. We start by doing some exploratory data analysis on the time series by decomposing it down into seasonal, trend and irregular components using moving averages in Appendix Figure \ref{fig:IntradayDecomp}. In addition, we plot the autocorrelation (ACF) in Figure \ref{fig:IntradayACF} and partial autocorrelation functions (PACF) in Figure \ref{fig:IntradayPACF}. From this, we find a strong seasonality component that repeats every 8 time steps (where one time step is one trading hour), exactly the length of market hours. The pattern of seasonality also exactly reflects our anecdotal observations, with an initial spike in volume at market open and a larger spike at the close.

We start by fitting SARIMA models with seasonality components with period 8. Our initial method to find the best SARIMA model for prediction was using the the R function \texttt{auto.arima()} at each timestep during cross validation. This function automatically finds the model with the lowest Akaike information Criterion (AIC) under the user-defined maximum order constraints. After running a full cross validation using this function, we examine which models with which orders were commonly fit. In fact only two models were ever found to be best, and these were ARIMA(1,0,3)$\times$(0,1,2)$_8$ and ARIMA(1,0,2)$\times$(0,1,2)$_8$. We hypothesize that using \texttt{auto.arima()} could lead to overfitting since it always maximizes model performance on the training set during cross validation. Thus, we run two more cross validation runs where we fix the model to either be ARIMA(1,0,3)$\times$(0,1,2)$_8$ or ARIMA(1,0,2)$\times$(0,1,2)$_8$. Comparing average MSE and MAPE values for these three cross validation experiments, our hypothesis was confirmed: the runs with fixed model types performed better as seen in Table \ref{tab:SARIMAmodeleval}. Interestingly ARIMA(1,0,3)$\times$(0,1,2)$_8$ has a lower average MSE score, but ARIMA(1,0,2)$\times$(0,1,2)$_8$ has a lower average MAPE score. But regardless the technique of fitting the best model at each timestep always leads to worse out-of-sample prediction performance, likely due to overfitting.

\begin{table}[h]
    \centering
    \begin{tabular}{|c|c|c|}
        \hline
        Model Choice & Average MSE & Average MAPE\\
        \hline
        Best model at each timestep & 8.060e+12 & 0.667\\
        $\textbf{ARIMA(1,0,3)$\times$(0,1,2)}_\textbf{8}$ & 7.902e+12 & \textbf{0.648}\\
        $\textbf{ARIMA(1,0,2)$\times$(0,1,2)}_\textbf{8}$ & \textbf{7.868e+12} & 0.674\\
        \hline
    \end{tabular}
    \caption{Model Selection for SARIMA (Intraday)}
    \label{tab:SARIMAmodeleval}
\end{table}


The next model we fit is a SARIMAX model with the previously defined set of technical indicators as our exogenous variables. We start with the null model and implement forward stepwise selection, adding variables that improve model fit as defined by cross-validation error. From Table \ref{tab:SARIMAXmodelevalIntraday}, we find that all models improve upon the SARIMA model with regards to average MSE. In addition, it appears to be optimal to add the three covariates Average Directional Index, Exponential Moving Average, and Momentum since the model with these covariates not only achieved the best average MSE but also the best average MAPE. We observe that adding any more covariates to this best model was detrimental to performance, likely because adding too many covariates allows the model to learn too many details in the training dataset which makes it harder to generalize to new data.

\begin{table}[h]
\parbox{0.55\linewidth}{
    \centering
    \begin{tabular}{|c|c|c|}
        \hline
        Covariates Used & Average MSE & Average MAPE\\
        \hline
        ADX & 7.716e+12 & 0.667\\
        ADX, EMA & 7.569e+12 & 0.724\\
        \textbf{ADX, EMA, MOM} & \textbf{7.416e+12} & \textbf{0.638}\\
        ADX, EMA, MOM, WPR & 7.439e+12 & 0.652\\
        \hline
    \end{tabular}
    \caption{Covariate Selection for SARIMAX (Intraday)}
    \label{tab:SARIMAXmodelevalIntraday}
}
\hfill
\parbox{0.5\linewidth}{
    \centering
    \begin{tabular}{|c|c|c|}
        \hline
        m & Average MSE & Average MAPE\\
        \hline
        2 & 1.046e+13 & 0.983\\
        \textbf{3} & \textbf{9.300e+12} & \textbf{0.644}\\
        5 & 1.090e+13 & 0.654\\
        10 & 1.135e+13 & 0.651\\
        25 & 1.289e+13 & 0.664\\
        100 & 1.726e+13 & 0.856\\
        \hline
    \end{tabular}
    \caption{m Selection for FDPR (Intraday)}
    \label{tab:MFreqIntraday}
}
\end{table}

Lastly, we investigate whether modeling our volume time series via FDPR would lead to any improvements. We plot the periodogram for our full data Appendix Figure \ref{fig:IntradayPeriodogram}, and it has only a few major spikes. This means we will likely not need a high value for the hyperparameter $m$ which controls how many sinusoidal functions are summed together to create our model. However, leakage does seem to be present in Figure \ref{fig:IntradayPeriodogram} with small spikes near large spikes, meaning our data may not contain Fourier frequencies and thus may not be represented well with sinusoidal functions with these frequencies. Cross validation results for various values of $m$ are shown in Table \ref{tab:MFreqIntraday}. Choosing $m$ to be 3 gives the best results, which makes sense given we only see a few spikes in the periodogram. Also, all FDPR models appear to perform worse than our SARIMAX ones, with average MSE values being significantly higher and average MAPE values being slightly higher. This highlights how this model does not have great predictive power, potentially since our data does not contain pure Fourier frequencies.

\begin{figure}[h]
    \centering
    \includegraphics[width=0.49\linewidth]{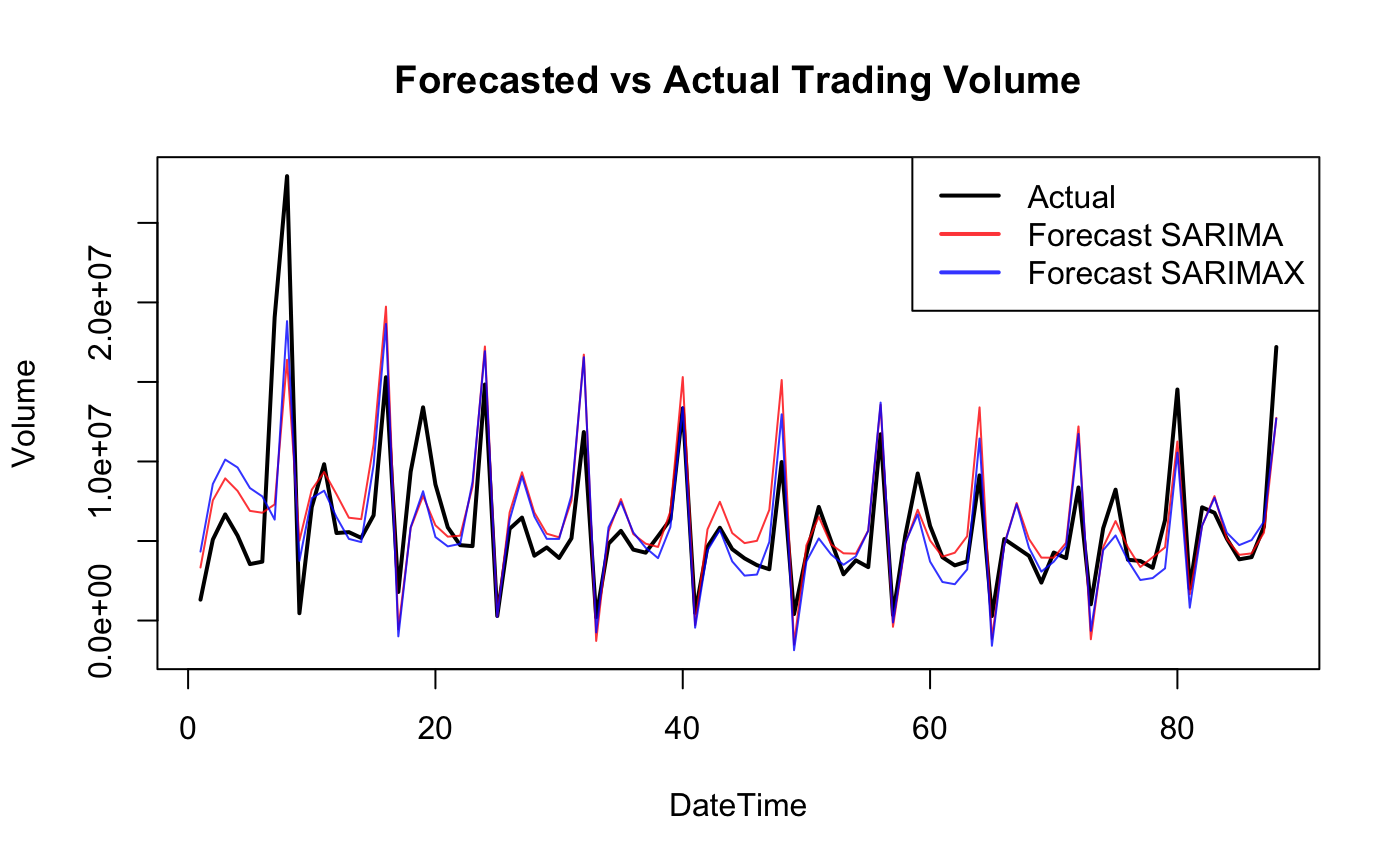}
    \includegraphics[width=0.49\linewidth]{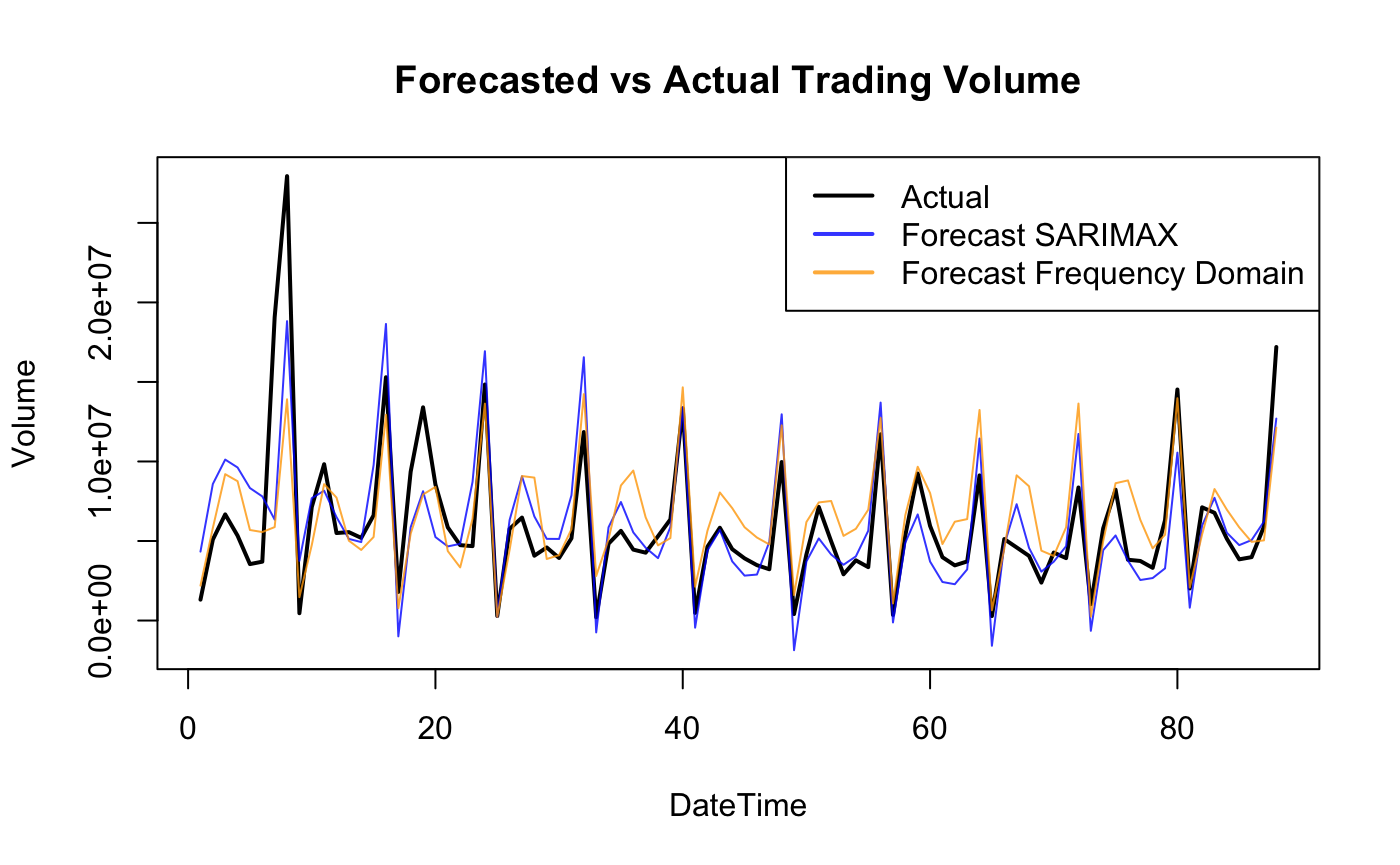}
    \caption{Forecasts for Intraday Data}
    \label{fig:IntradayForecast}
\end{figure}

However, looking at a few metrics to asses model performance can be limiting. Thus, we plot our predicted values for our various models to better assess predictive power. We start by comparing SARIMA and SARIMAX in the left-hand graph of Figure \ref{fig:IntradayForecast}. We can see that many red peaks are above blue peaks, meaning the SARIMA model tends to over-predict volumes for periods of time that experience lots of trades. Also at low points the blue SARIMAX line seems closer to the black actual volume line, indicating the superior performance of SARIMAX for low volume times too. Since SARIMAX has lower average MSE and MAPE values as well as a better visual fit, we moved on to comparing its predictions with those of the FDPR. In the right-hand side of \ref{fig:IntradayForecast} we see that for certain times the blue peaks exceed the yellow peaks, but for others the reverse is true. Thus, it is unclear which model performs better for time periods where the actual volume is high. But looking at low points in actual volume, it is obvious that SARIMAX prevails: the yellow lines tend to be quite a bit higher than the actual black volume values.

\begin{figure}[h]
\includegraphics[width=10cm]{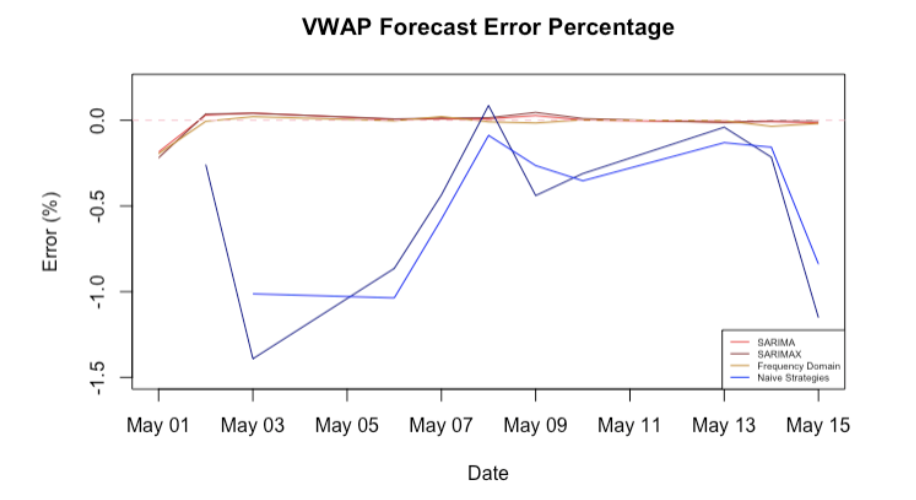}
\centering
\caption{Errors in VWAP Predictions of Our Models versus Naive Baselines (Intraday)}
\label{VWAPIntraday}
\end{figure}

Next we evaluate the VWAP error of our best SARIMA, SARIMAX, and frequency domain models and compare them to naive baselines as this is a metric used frequently by traders. Our two baselines are first a VWAP model that predicts that there is no change between yesterday and today's VWAP and second a VWAP prediction based off a 3-day rolling average. As shown in Figure \ref{VWAPIntraday}, we find that our models, shown in red, perform significantly better than both these baselines, shown in blue, with errors within 0-5 basis points, excluding that of the first day, whereas the VWAP error of the baselines are much more volatile and reach values close to 1.5\%. Note that for simplicity all VWAP estimates assume perfect price forecasting.

\subsection{Daily}
Although we have a good prediction model for intraday data using SARIMAX, daily data can look very different. But it is just as important to be able to predict daily volume data so traders can plan out multi-day strategies (e.g. entering or liquidating large positions). For the intraday data, we predict the values of the following day, or following 8 trading hours, in our cross validation loops. Here again we predict one day in the future, something which is commonly done since stock volumes can be very volatile and environments can change quickly.

Again we begin with EDA. In the Appendix we plot the ACF (Figure \ref{fig:acf_dly}) and PACF (Figure \ref{fig:pacf_dly}) for the entire daily dataset. Here there is no clear seasonality present and instead an ARIMA model seems like the best first step. This is because the ACF values seem to decrease exponentially, and the PACF values are only above the threshold for a few lags. Proceeding similarly to the intraday analysis, we perform cross validation at first by using the \texttt{auto.arima()} function to fit the best possible model at each timestep according to AIC values. But unlike with the daily data, we have that \texttt{auto.arima()} always fits the exact same model every time, namely an ARIMA(3,1,2) model. This model achieves an average MSE of 3.307e+14 and an average MAPE of 0.210. As this seems to be the best possible performance using an ARIMA model, we move onto more complicated models right away.

We again consider adding covariates to our model. But since seasonality does not seem to be present in our daily data, we consider ARIMAX models. This is identical to SARIMAX models but just removes seasonal autoregressive, moving average, and differencing terms. Through the use of exogneous variables, SARIMAX can potentially forecast better than our SARIMA models that did not have covariates. The same method was used to find the most appropriate covariates. We start with a null model and add one of the six covariates defined in Appendix Table \ref{tab:TechnicalIndicators} that achieves the best performance, and then continue adding covariates one by one until overfitting is noticed. As seen in Table \ref{tab:ARIMAXmodelevalDaily} overfitting occurs very quickly, or after adding just the second covariate. This means that the best possible ARIMAX model just includes the momentum technical indicator and also shows significant improvements on both metrics over the best ARIMA model found above.
 
\begin{table}[h]
\parbox{0.5\linewidth}{
    \centering
    \begin{tabular}{|c|c|c|}
        \hline
        Covariates Used & Average MSE & Average MAPE\\
        \hline
        \textbf{MOM} & \textbf{2.693e+14} & \textbf{0.198}\\
        MOM, RSI & 2.846e+14 & 0.206\\
        MOM, RSI, ADX & 2.814e+14 & 0.209\\
        \hline
    \end{tabular}
    \caption{Covariate Selection for ARIMAX (Daily)}
    \label{tab:ARIMAXmodelevalDaily}
}
\hfill
\parbox{0.55\linewidth}{
    \centering
    \begin{tabular}{|c|c|c|}
        \hline
        m & Average MSE & Average MAPE\\
        \hline
        2 & 3.350e+14 & 0.230\\
        3 & 3.122e+14 & 0.209\\
        5 & 3.552e+14 & 0.205\\
        10 & 3.179e+14 & 0.192\\
        25 & 2.922e+14 & 0.206\\
        \textbf{40} & \textbf{2.773e+14} & \textbf{0.197}\\
        50 & 2.862e+14 & 0.208\\
        100 & 2.970e+14 & 0.214\\
        \hline
    \end{tabular}
    \caption{m Selection for FDPR (Daily)}
    \label{tab:MFreqDaily}
}
\end{table}

Lastly, we make an attempt to find an FDPR for our daily volume time series. Looking at the periodogram of our full daily data in Appendix Figure \ref{fig:DailyPeriodogram}, we see a very different picture than with the intraday data. There are no obvious spikes in the graph, and many frequencies seem to be somewhat important in the time series. Thus, we expect a much higher value for the hyperparameter $m$, and this is confirmed by our results. In Table \ref{tab:MFreqDaily} we see that the optimal $m$ value is 40, much higher than the value of 3 we find for intraday. But unlike the intraday data, the average MSE is extremely close to our best model so far and the average MAPE is actually the lowest seen for daily data. Since these metrics are relatively close for our best ARIMA, ARIMAX, and FDPR models, we visualize our predictions to better distinguish between their performances.

\begin{figure}[h]
    \centering
    \includegraphics[width=0.49\linewidth]{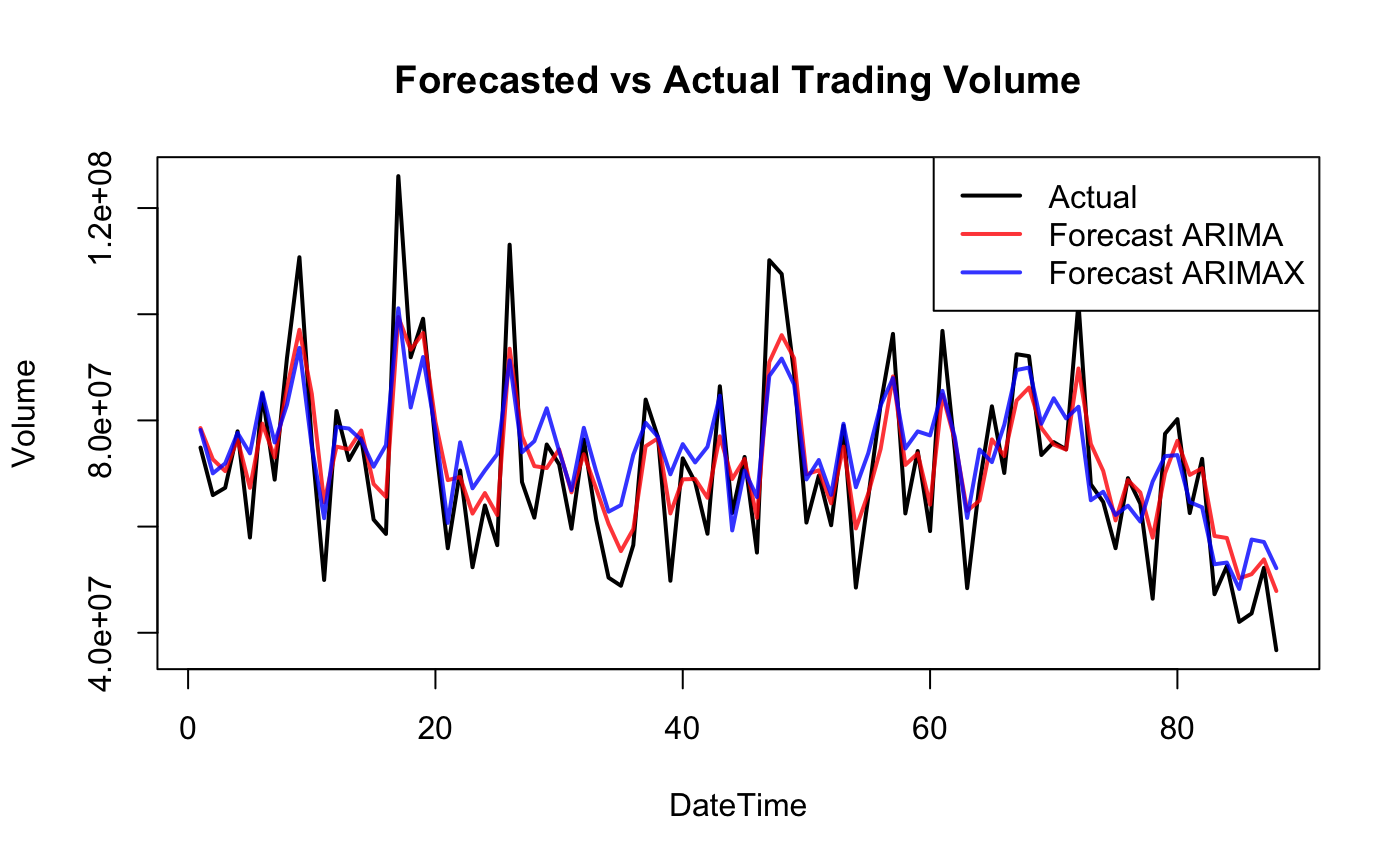}
    \includegraphics[width=0.49\linewidth]{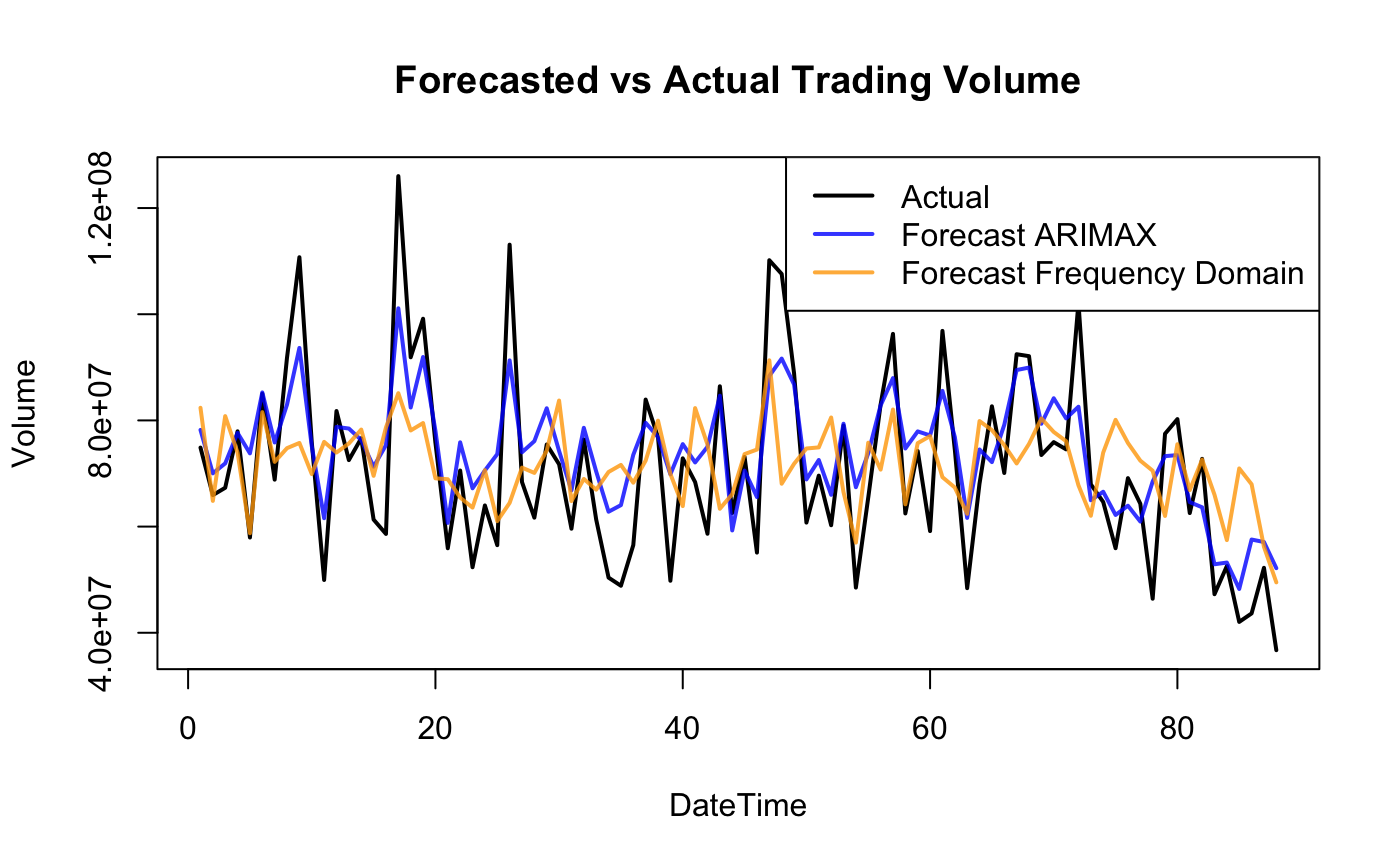}
    \caption{Forecasts for Daily Data}
    \label{fig:DailyForecast}
\end{figure}

First looking at the left-hand plot of Figure \ref{fig:DailyForecast}, we see that ARIMA and ARIMAX give relatively similar predictions, and both are extremely close to the true volume values in black. Both high and low points in volume are traced very closely, and since it is hard to distinguish the performance between these two values visually, we proceed with ARIMAX since its average MSE and MAPE were lower. Then shifting our focus to the right-hand side of Figure \ref{fig:DailyForecast}, the FDPR does seem to have significantly worse predictions than ARIMAX. It estimates the mean volume decently but cannot anticipate fluctuations as well. Rather, the yellow line seems to stay relatively flat in between large high and low points in volume. Thus, the FDPR likely has less flexibility than our ARIMA or ARIMAX models.

\begin{figure}[htp]
    \centering
    \includegraphics[width=9cm]{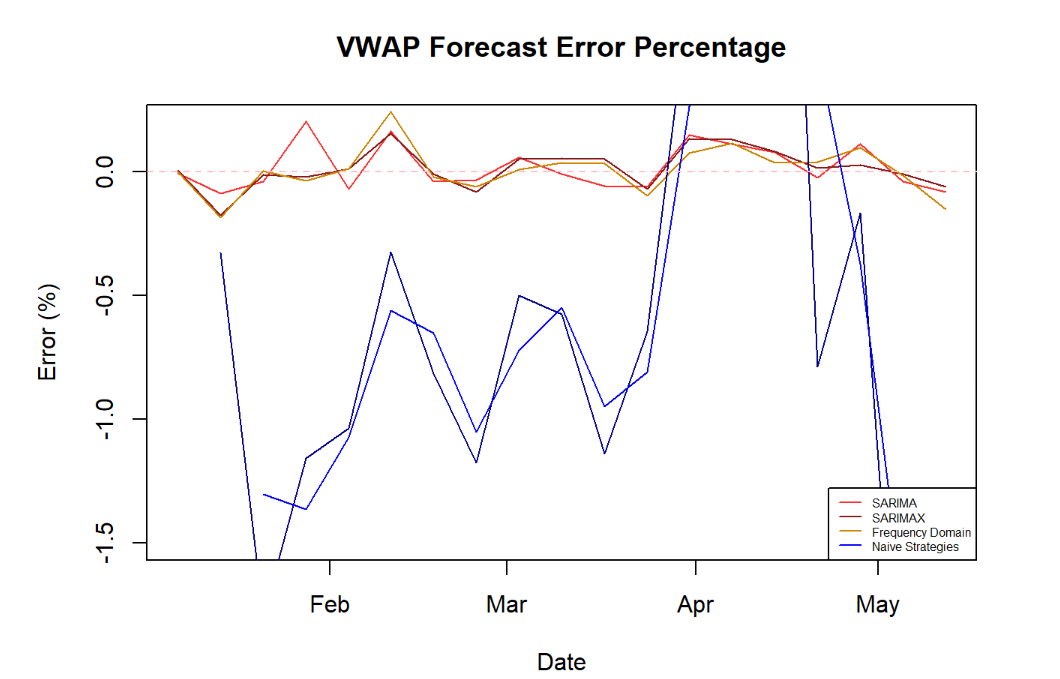}
    \caption{Errors in VWAP Predictions of Our Models versus Naive Baselines (Daily)}
    \label{fig:VWAPdaily}
\end{figure}

We then plot the VWAP error for our best SARIMA, SARIMAX, and spectral representation models for our daily SPY data. VWAP calculations require a range of time, but we only predicted single values at each timestep. Thus, we chose to group predictions by week in order to compare predicted versus realized VWAP. Figure \ref{fig:VWAPdaily} reveals our models perform significantly better than the naive baselines. The VWAP for our models remain tight to realized VWAP whereas the error of the baselines are more volatile and much larger in magnitude. However, the predicted values here for daily do seem to deviate more compared to those for intraday, signifying our daily models performed more poorly overall.

\section{Conclusion}

In this analysis, we focus on maximizing predictive power of time-series models in forecasting the intraday and daily trading volume of SPY according to the metrics MSE, MAPE, and VWAP error. Our intraday analysis indicates that using SARIMAX with the exogenous variables average directional index, exponential moving average, and momentum give us the optimal forecast, outperforming SARIMA and a spectral representation of the data using $m = 3$ Fourier frequencies. However, all three models significantly outperform our naive baselines with respect to tracking VWAP. However, when performing our analysis of daily volume data, we see a lack of seasonality. This is confirmed by R choosing ARIMA and ARIMAX models over SARIMA and SARIMAX models during cross validation. Also, a higher $m$ value of 40 for the FDPR model is optimal. But using exogenous variables with ARIMAX still gives the best predictions for daily data. Overall, we have shown that trading volume can be accurately predicted using ARIMA models with exogenous variables and adding seasonal components when necessary.

In this paper we restricted ourselves to two modeling methods: ARIMA-based methods and FDPR representations. However, related work has found success in using other approaches such as nonlinear threshold time-series models or neural based methods such as LSTMs. Extensions of this analysis may also incorporate deep learning models to help improve forecasting performance. Additionally, we restrict ourselves to a set of 6 technical indicators. A more comprehensive study may include additional exogenous covariates that build up the number of technical indicators, as well as include additional economic indicators that are known to influence the markets. Finally, VWAP-based trading strategies are only one of many use-cases of forecasting volume. In addition to VWAP error, another useful evaluation metric would be to use the volume forecasts as an input into alpha generating trading strategies.

\newpage

\bibliographystyle{unsrtnat}
\bibliography{references}

\newpage
\section{Appendix}
\renewcommand{\thefigure}{A\arabic{figure}}
\setcounter{figure}{0}
\renewcommand{\thetable}{A\arabic{table}}
\setcounter{table}{0}
SARIMA Model Formula for process $Y_t$ and white noise $Z_t$:
\begin{equation}
    \Phi(B^s)\phi(B)\triangledown_s^D\triangledown^d Y_t=\delta+\Theta(B^s)\theta(B)Z_t
\label{eq:SARIMAModel}
\end{equation}

SARIMAX Model Formula for process $Y_t$, white noise $Z_t$, and covariates $X_{i_t}$, which refers to the $i$th covariate in time period $t$:
\begin{equation}
    \Phi(B^s)\phi(B)\triangledown_s^D\triangledown^d Y_t=\delta+\Theta(B^s)\theta(B)Z_t+\sum_{i=1}^p\beta_i X_{i_t}
\label{eq:SARIMAXModel}
\end{equation}
where $\Phi(B^s)$ is the seasonal AR operator, $\phi(B)$ the AR operator, $\Theta(B^s)$ the seasonal MA operator, $\theta(B)$ the MA operator, $\triangledown_s^D$ the seasonal differencing operator, $\triangledown^d$ the differencing operator, and $\delta$ a parameter related to drift. We also suppose that $Z_t$ is a white noise process.

\begin{table}[h]
    \centering
    \begin{tabular}{|c|c|c|}
        \hline
        Technical Indicator & Abbreviation & Description\\
        \hline
        Average Directional Index & ADI & strength or weakness of observed trend in market\\
        Exponential Moving Average & EMA & mean of previous closing prices, weighted towards recent ones\\
        Momentum & MOM & difference between current and past closing prices\\
        Rate of Change & ROC & percent change in current versus past closing prices\\
        Relative Strength Index & RSI & weakness of observed trend given previous closing prices\\
        Williams \%R & WPR & measures if stock has been over-bought or -sold\\
        \hline
    \end{tabular}
    \caption{Technical Indicators considered for SARIMAX}
    \label{tab:TechnicalIndicators}
\end{table}

\begin{figure}[htp]
    \centering
    \includegraphics[width=9cm]{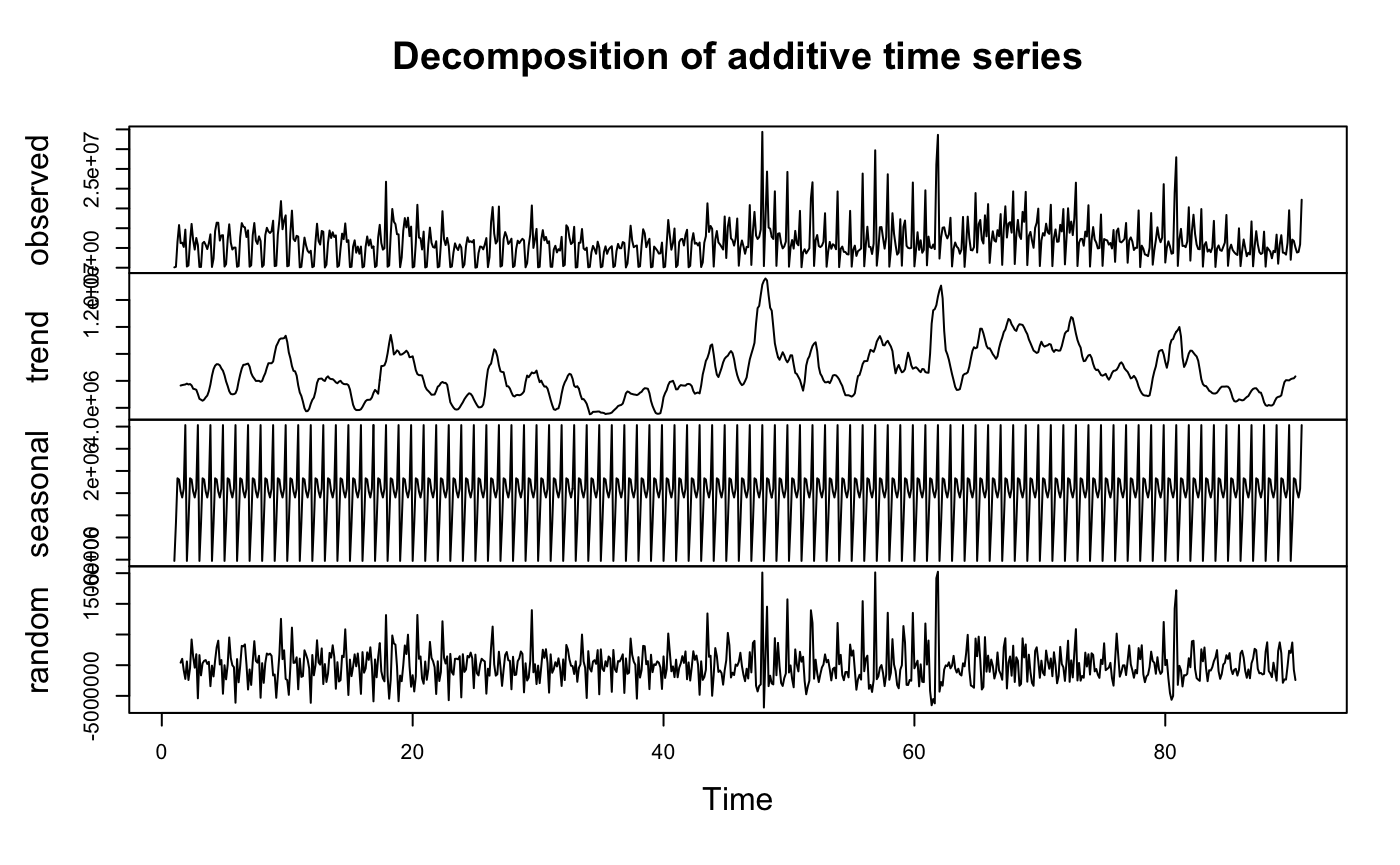}
    \caption{Intraday Decomposition}
    \label{fig:IntradayDecomp}
\end{figure}

\begin{figure}[htp]
    \centering
    \includegraphics[width=9cm]{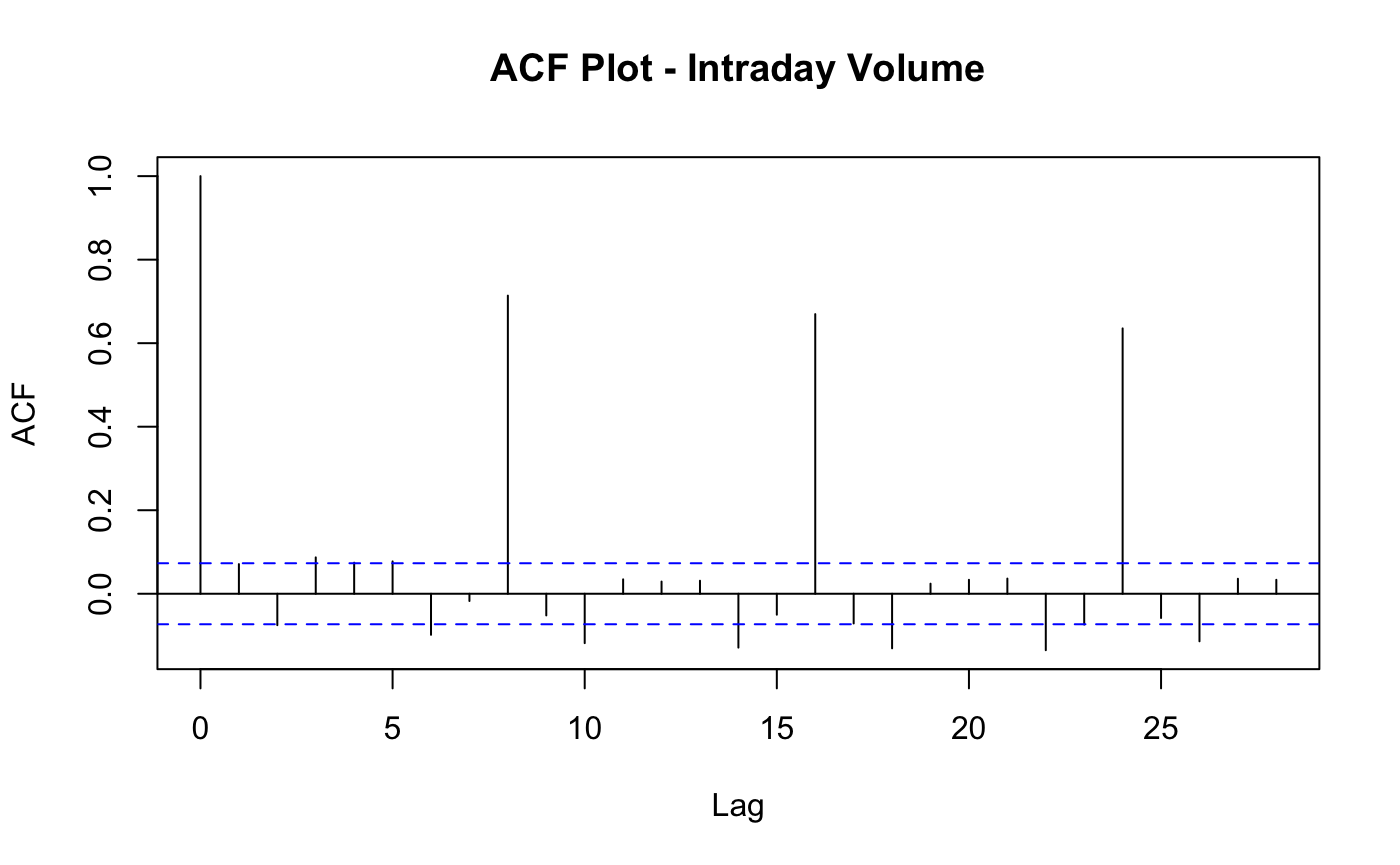}
    \caption{Intraday ACF}
    \label{fig:IntradayACF}
\end{figure}

\begin{figure}[htp]
    \centering
    \includegraphics[width=9cm]{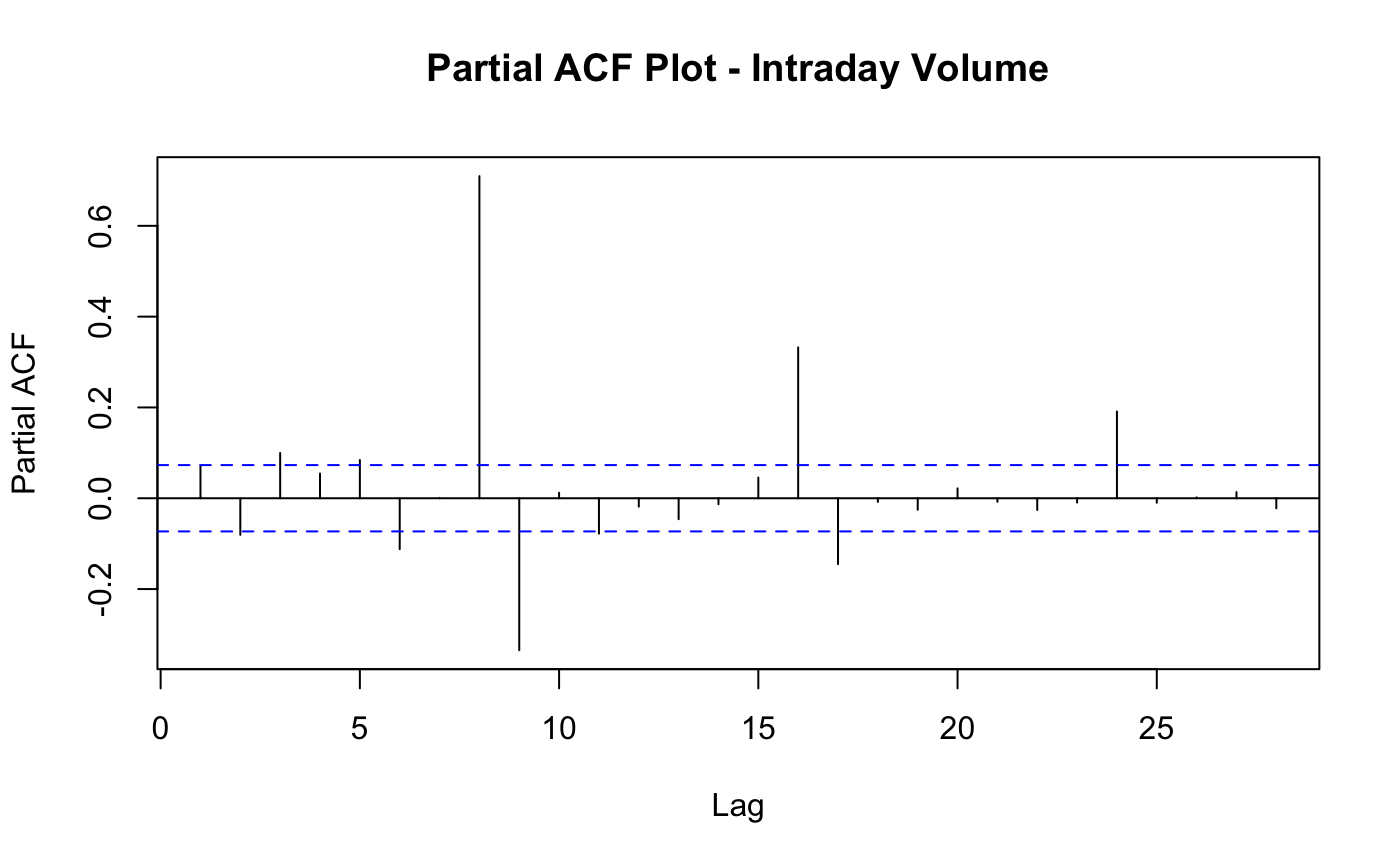}
    \caption{Intraday PACF}
    \label{fig:IntradayPACF}
\end{figure}

\begin{figure}[htp]
    \centering
    \includegraphics[width=9cm]{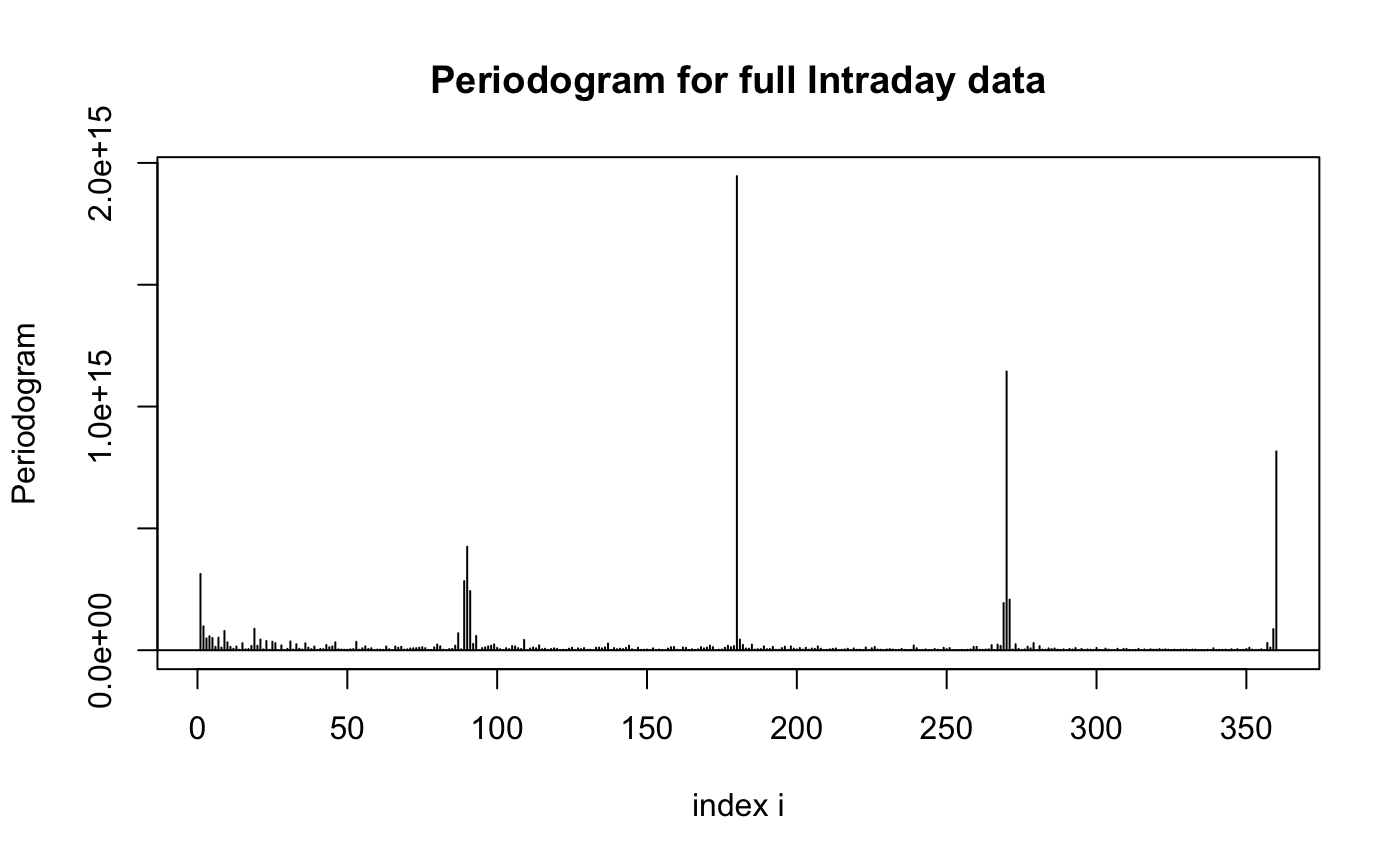}
    \caption{Intraday Periodogram}
    \label{fig:IntradayPeriodogram}
\end{figure}

\begin{figure}[htp]
    \centering
    \includegraphics[width=9cm]{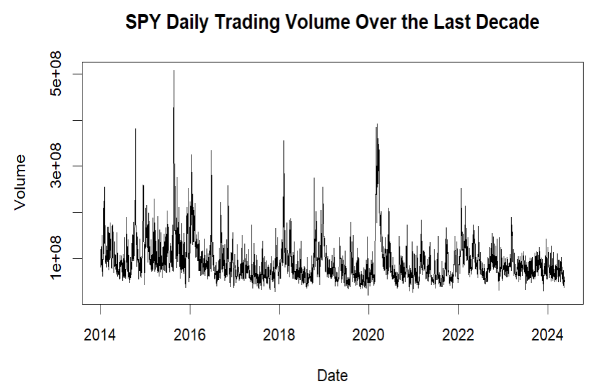}
    \caption{Daily Overview}
    \label{fig:spydcd}
\end{figure}

\begin{figure}[htp]
    \centering
    \includegraphics[width=9cm]{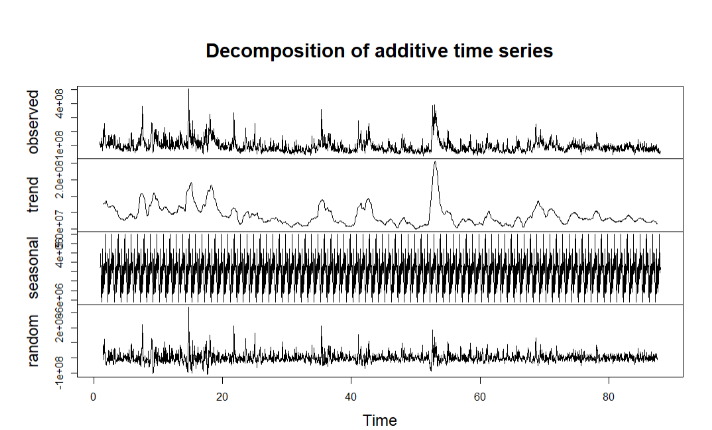}
    \caption{Daily Decomposition}
    \label{fig:spydcmp}
\end{figure}

\begin{figure}[htp]
    \centering
    \includegraphics[width=9cm]{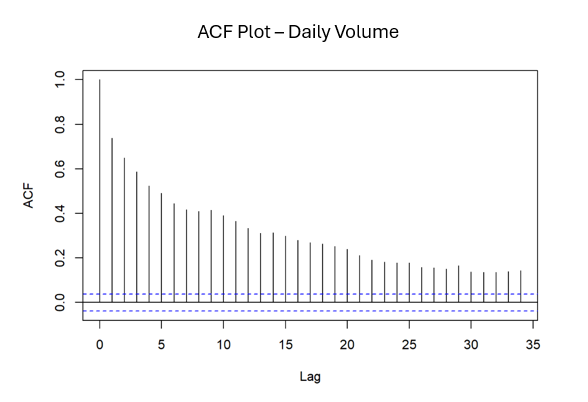}
    \caption{ACF: Daily Volume}
    \label{fig:acf_dly}
\end{figure}

\begin{figure}[htp]
    \centering
    \includegraphics[width=9cm]{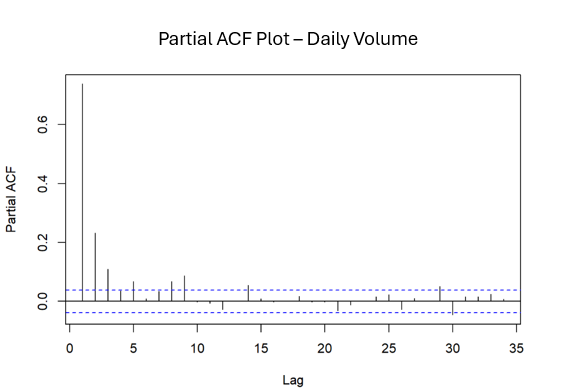}
    \caption{PACF: Daily Volume}
    \label{fig:pacf_dly}
\end{figure}

\begin{figure}[htp]
    \centering
    \includegraphics[width=9cm]{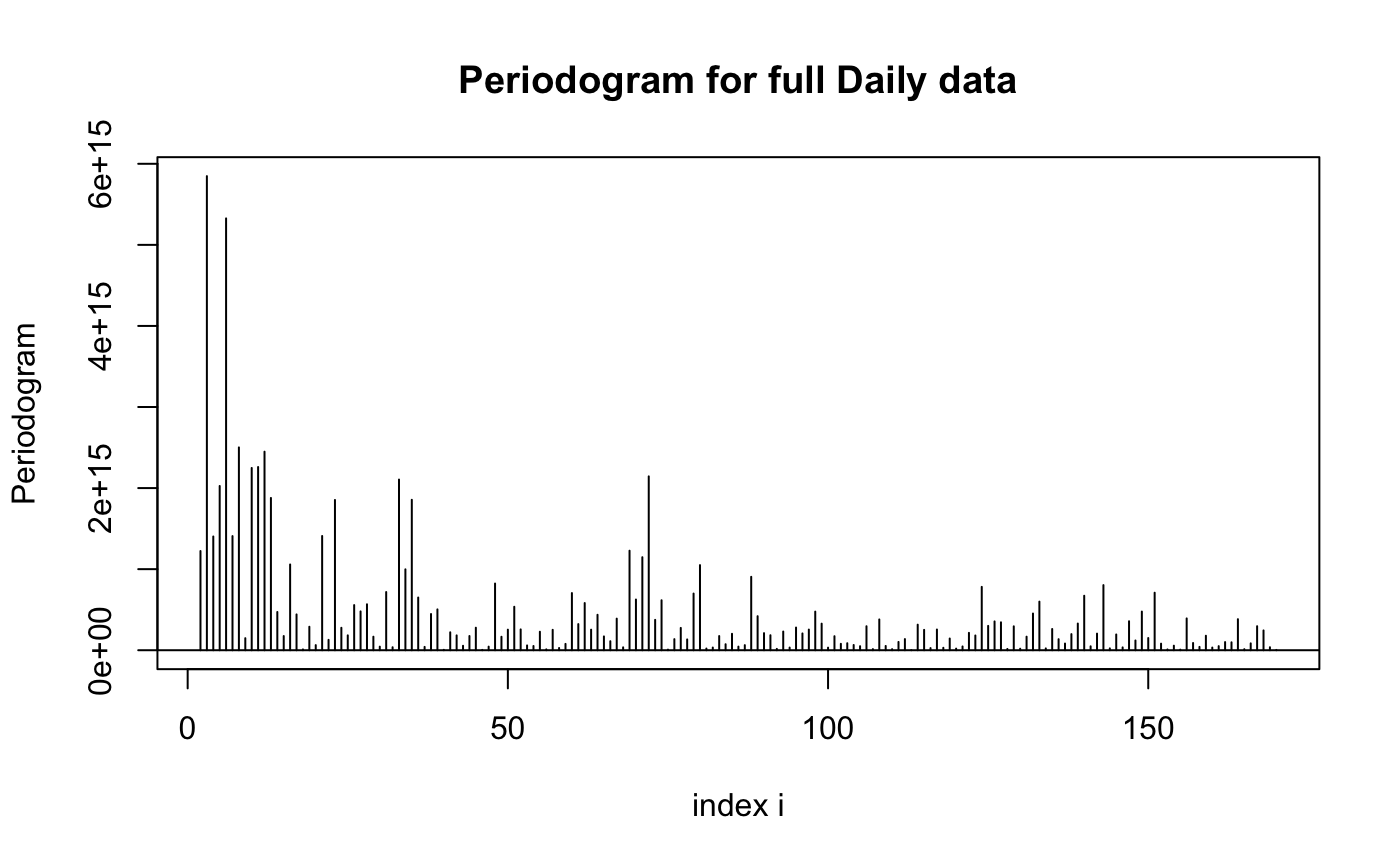}
    \caption{Daily Periodogram}
    \label{fig:DailyPeriodogram}
\end{figure}

\end{document}